\documentclass[twocolumn,floats,,amsmath,amssymb,superscriptaddress]{revtex4}

\usepackage{graphicx}
\usepackage{dcolumn}
\usepackage{bm}

\begin{document}

\title{Structural and electronic properties of pentacene molecule and molecular pentacene solid}

\author{R.~G.~Endres}
\affiliation{Center for Computational Sciences and Computer Science \& Mathematics Division, 
Oak Ridge National Laboratory, Oak Ridge, TN 37831-6114 USA}

\author{C.~Y.~Fong}
\affiliation{Department of Physics, University of California, Davis, 
             CA 95616-8677 USA}

\author{L.~H.~Yang}
\affiliation{H Division, Lawrence Livermore National Laboratory, 
             Livermore, CA 94551 USA}

\author{G. Witte}
\affiliation{Lehrstuhl f\"ur Physikalische Chemie I, 
    Ruhr-Universit\"at Bochum, Universit\"atsstrasse 150, 44780 Bochum, Germany}

\author{Ch.~W\"oll}
\affiliation{Lehrstuhl f\"ur Physikalische Chemie I, 
    Ruhr-Universit\"at Bochum, Universit\"atsstrasse 150, 44780 Bochum, Germany}

\date{\today}

\begin{abstract}
The structural and electronic properties of a single pentacene molecule and a 
pentacene molecular crystal, an organic semiconductor, are examined by a first-principles method
based on the generalized gradient approximation of 
density functional theory.
Calculations were carried out for a 
triclinic unit cell containing two pentacene molecules. The bandwidths of the 
valence and conduction bands which determine the charge migration mechanism are 
found to depend strongly on the crystallographic direction.
Along the triclinic reciprocal lattice vectors {\bf A} and {\bf B} which are orientated
approximately  perpendicular to the molecular axes the maximal valence (conduction) band width amounts to
only 75 (59) meV, even smaller values are obtained for the {\bf C} direction 
parallel to molecular axes even less.
Along the stacking directions {\bf A+B} and {\bf A-B}, however, 
the maximal valence (conduction) band width is found to reach 145 ($260$) meV. 
The value for the conduction band width is larger than estimates 
for the polaron binding energy but significantly smaller than recent
results obtained by semiempirical methods.
The single molecule has a HOMO-LUMO gap of about 1.1 eV as deduced from the Kohn-Sham eigenvalue 
differences. When using the self-consistent field method, which is expected to yield more reliable results,
a value of 1.64 eV is obtained. The theoretical value for the band gap in the molecular solid
amounts to 1.0 eV at the $\Gamma$-point.

\end{abstract}

\maketitle

\section{\label{sec1}Introduction}

Organic semiconductors based on pentacene ($C_{22}H_{14}$) or
other aromatic hydrocarbon molecules have recently attracted an
enormous interest regarding their use for molecular electronics
\cite{horowitz}. A pentacene field effect transistor showing
a temperature-independent high-mobility ($\mu\sim 1.3 cm^2/Vs$) 
has been fabricated\cite{nelson}. It has to be noted, however, that in the latter case 
polycrystalline material was used; considerable higher mobilities can be expected for single crystalline material.
Although mobilities for single crystalline pentacene have not yet been measured, recent
work has demonstrated that epitaxial growth of pentacene on solid substrates may be possible
\cite{lukas}, thus putting the realization of devices with an active layer of single-crystalline pentacene
into sight. 
Despite this pronounced interest, however, there are still many open questions even on rather basis properties 
on molecular semiconductors like pentacene, in particular as far as the electronic structure is concerned.
The conventional view is that in molecular crystals of aromatic molecules like pentacene
hole transport is dominant and that 
the valence band width is of order 100 meV - much smaller than the 
experimental 500 meV estimate\cite{heeger} for the conducting charge transfer salt 
TTF-TCNQ (tetrathiafulvalene-tetracyanoquino dimethane).
The bandwidth is further important for understanding the charge transport
mechanism. According to Holstein's polaron model\cite{holstein}, the small polaron
limit (size of lattice constant) is reached when the electronic bandwidth
is small compared to the polaron binding energy. The binding energy
is experimentally estimated to be of order $\sim 200$ meV\cite{brown}. 
On the theoretical side, however, there have only been very few reports. Using semiempirical methods,
results ranging between  $\sim 600 meV$ obtained using the cluster approach
with a semiempirical Hartree-Fock INDO (intermediate neglect of differential
overlap)\cite{cornil} and  $\sim 120 meV$ obtained from an 
extended-H\"uckel-type (EHT)-calculation\cite{haddon} were reported.

In this paper, we address the above mentioned issues 
by using a first-principles algorithm \cite{siesta_code}. In particular, we
illustrate how the bandwidths of the VB and CB depend on the orientation of
the herringbone structured crystal. We find that although there is only weak dispersion of the bands
along the reciprocal lattice vectors {\bf A}, {\bf B}, and especially {\bf C} (along the molecules), 
along the stacking directions {\bf A+B} and {\bf A-B} the dispersion reaches significantly larger values. 
For these directions, 
the bandwidth of the CB and the next higher band CB+1 amounts to $149$ and $260$ meV. For the {\bf A-B}
direction the VB and the next lower band VB-1 don't become quite as large,
the corresponding values amount to $145$ and $131$ meV. 

Hence, the band widths in the stacking directions can become larger 
than estimates for the polaron binding energy. This should make
band-like transport at low temperatures possible. 
To the best of our knowledge this is the first fully {\it ab initio} bandstructure
calculation and theoretical band gap estimate in the literature. 

In section \ref{sec2}, we discuss briefly the method of calculation. It will be 
followed by a section describing the models in section \ref{sec3}. 
Results of the single molecule and the molecular solid and 
discussions will be presented in section \ref{sec4}. 
Finally, in section \ref{sec5}, a summary will be given.

\begin{figure}
\includegraphics[width=8.5cm,angle=0]{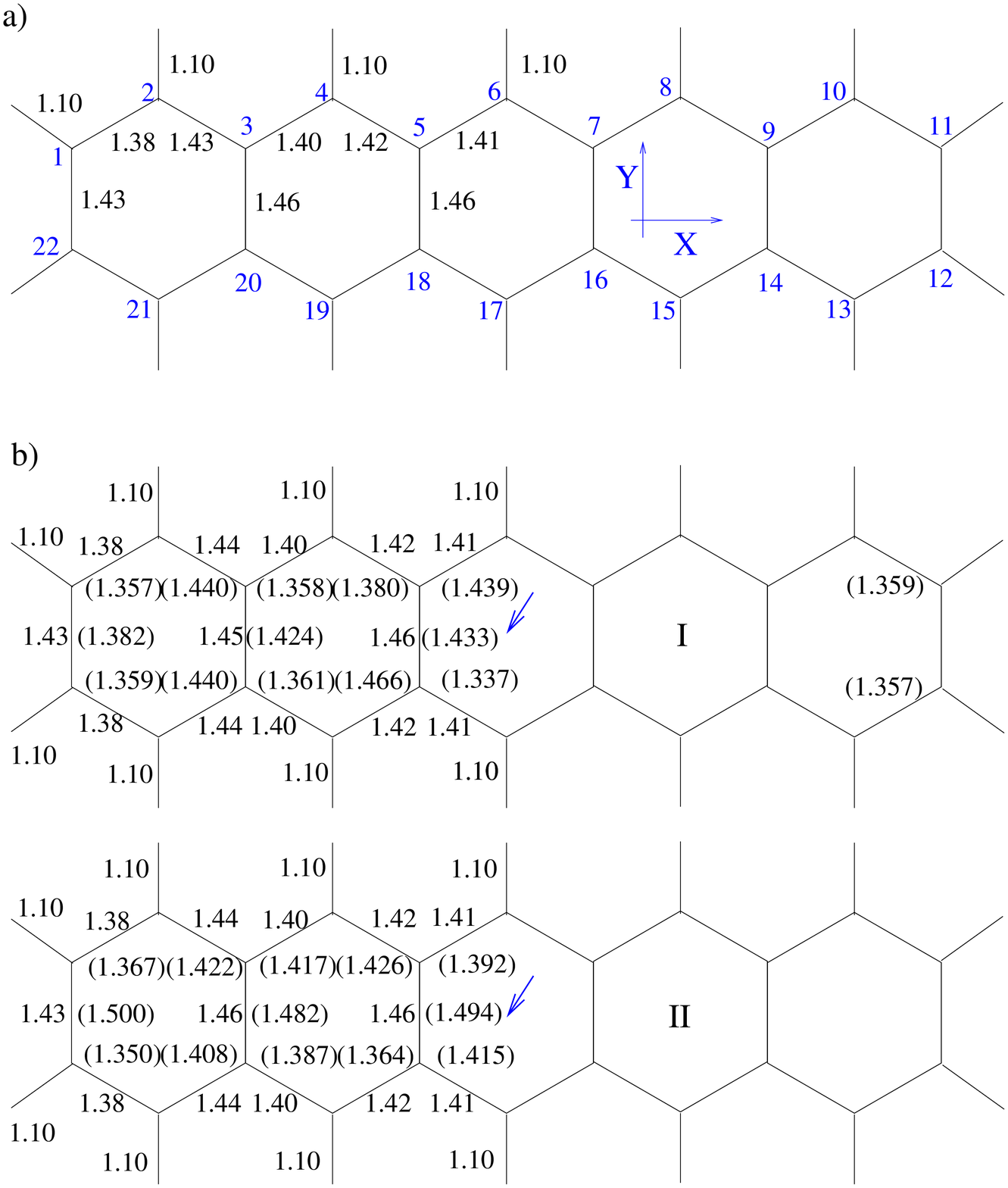}\\
\caption{\label{fig:pos}Atomic bond-lengths a) of a fully relaxed
single pentacene molecule (in xy-plane) and b) of the 
constrained relaxed pentacene crystal. 
One can clearly see the alternating double (shorter) and 
single (longer) bonds. The experimental values\cite{campbell} are shown in brackets (...). 
The missing bond-lengths follow from symmetry.}
\end{figure}

\section{\label{sec2}Methods of calculation}
We used one of the popular density functional theory (DFT)
algorithms with localized orbitals as the basis functions,
SIESTA\cite{siesta_papers}. It uses Troullier-Martins
norm-conserved pseudopotentials\cite{TM} in the Kleinman-Bylander
separable form\cite{KB}. The basis set is made of pseudo atomic
orbitals (PAO) of multiple-zeta form including optionally
polarization orbitals. The first-zeta orbitals are obtained by the
method of Sankey and Niklewski\cite{sankey}, while the second-zeta
orbitals are constructed in the split-valency philosophy well
known from quantum chemistry\cite{gaussian}. With this basis set,
SIESTA calculates the self-consistent potential on a grid in real
space. The fineness of this grid is determined in terms of an
energy cutoff $E_c$ in analogy to the energy cutoff when the basis
set involves plane waves. In the present calculations, we used
$E_c$ to be 80 Ry and a double-zeta plus polarization orbitals
(DZP) basis set. For the exchange-correlation energy functional,
the generalized gradient approximation (GGA) in the version of
Perdew-Burke-Ernzerhof\cite{PBE} is applied for characterizing
semi-nonlocal effects.

\section{\label{sec3}Models}

\subsection{\label{subsec3.1}Single molecule}

For the case of simulating a single molecule, we used a large
cubic supercell ($50\AA\times 50\AA\times 50\AA$) and placed the molecule at the center. 
These dimensions are much larger then the pentacene dimensions
($14.21\AA \times 5.04\AA$ in x-y plane) and results in no interactions between the unit cells.
The atoms in the molecule were relaxed until the magnitude of the force on each
atom is less than the tolerance $0.04 eV/\AA$.

\subsection{\label{subsec3.2}Molecular solid}

For the molecular solid, we used a triclinic unit cell containing
two non-equivalent molecules. The unit cell data with the Bravais
lattice vectors {\bf a}, {\bf b} (perpendicular molecular axes),
and {\bf c} (along molecular axes) and atomic positions were taken
from a previous x-ray measurement on a trichlorobenzene 
solution-gown single crystals
by R. Campbell and co-workers (1961)\cite{campbell}. The lattice vectors are 
$a=7.93$, $b=6.14$, $c=16.03\AA$, 
$\alpha=101.9$, $\beta=112.6$, $\gamma=5.8^o$.
There are other structure determinations, e.g. by 
D. Holmes {\it et al.} (1999)\cite{holmes}.
Although the crystallization solvent was the same in both cases,
the unit cell parameters disagree slightly. Their 
structure is similar to crystals obtained from the
vapor-phase deposition technique which results in
slightly denser molecular packing (unit cell volume reduced by 3\%)\cite{siegrist}. 
Polymorphism is common in molecular crystals and depends on
the applied preparation techniques and conditions\cite{siegrist}.
The main structural features however are the same
and we restrict ourselves to the analysis of pentacene crystal described
by Campbell's data.

Since the GGA functional does not include van der
Waals attraction\cite{adamo,kafafi},
we relaxed the atoms except 3 in each molecule in order to fix the
planes of the molecules. The force tolerance is again $0.04 eV/\AA$.
The {\bf k}-point sampling is done
just using the $\Gamma$-point. This is expected to be sufficient
due to the large unit cell.

In {\bf k}-space, we denote the reciprocal lattice vectors {\bf
A}, {\bf B}, and {\bf C}. We note however that the unit cell is
triclinic and not tetragonal, although the unit cell angles are
close to $90^o$. Hence by convention, e.g. {\bf A} is parallel to
$\bm b\times \bm c$ but not to {\bf a}.

\begin{figure}
\hspace{-2.0cm}
\begin{minipage}{8.5cm}
\includegraphics[width=10.0cm,angle=0]{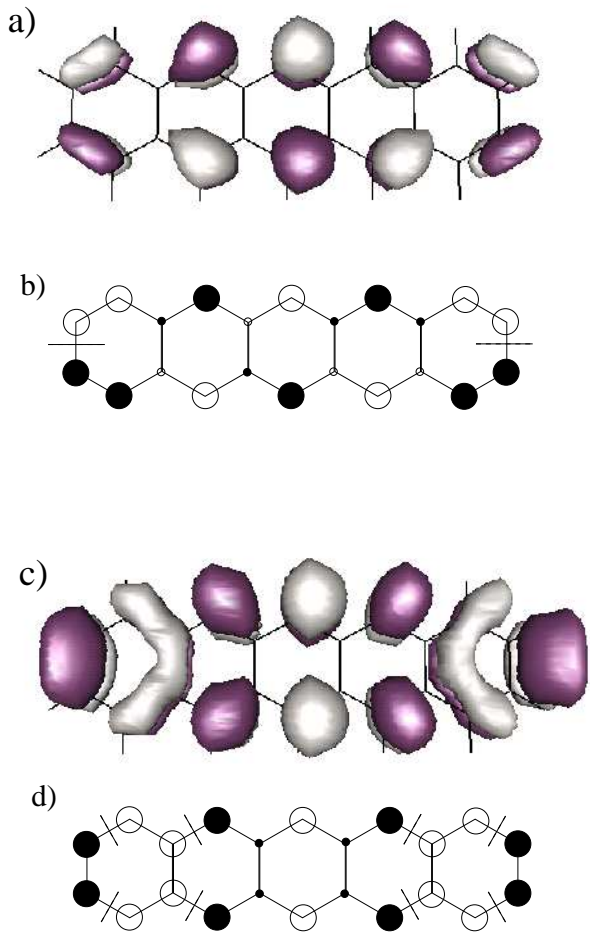}
\end{minipage}
\caption{\label{fig:wf}Single particle wavefunctions of isolated pentacene molecule.
The dark color represents a positive sign, the light color a negative sign.
(a) and (b) show the HOMO, (c) and (d) the LUMO. (a) isosurface at $\pm 0.05/\AA^{3/2}$; (b)
LCAO coefficients: small circles $0.01$ - $0.09$, large circles $0.12$ - $0.3$,
the short lines show regions of high gradients; (c) isosurface at $\pm 0.03 /\AA^{3/2}$, (d)
similar to (b).  Wavefunctions plotted with
the program gOpenMol\cite{gopenmol}.}
\end{figure}

\section{\label{sec4}Results and discussions} 

\subsection{\label{subsec4.1}Single molecule} 

In figure \ref{fig:pos} a) we show the relaxed single molecule. We
label the atoms and give the bond-lengths. The {\bf z}-coordinates for the 
relaxed atoms are less than $0.67 m\AA$ about $z=0$ showing that the molecule 
is essentially planar, in agreement with the 
experimental findings~\cite{holmes}.
The molecule is symmetric with respect to 
the C atoms 6 and 17 and the midpoints of the bonds 1-22 and 11-12.   
The bond-length in the {\bf y}-direction is typically $1.46\AA$, 
except at the ends where it is $1.43\AA$. 
We will compare the C-C and C-H bond-lengths to experimental crystallographic data  
in section \ref{subsec4.2}. In general they are similar to those in graphite ($1.42\AA$).
Our results show also smaller bond-lengths for the C atoms near the ends (bonds 
1-2, 10-11, 12-13, and 21-22) to their respective neighbors along the 
{\bf x}-direction. These bond-lengths are $1.38\AA$. All the C-H bonds
are $1.10\AA$ long.

The energy gap $\Delta$ separating the HOMO and the LUMO states is
$1.1$ eV as deduced from Kohn-Sham eigenvalue differences. This value
is about 40\% smaller than the experimental value, $1.82$ eV,
reported from ellipsometric spectra of thin pentacene films
\cite{park}. This underestimation of gaps by this magnitude is a
common deficiency of the GGA functional \cite{gap_probl}. Since
DFT is a groundstate theory, better results can usually be
obtained from total energy differences (self-consistent field
method)\cite{SCF}
\begin{equation}
\Delta = E_0(N) - E_0(N+1),\label{gap}
\end{equation}
where $E_0(N)$ is the groundstate of the neutral N (even) electron system
and $E_0(N+1)$ is the groundstate with one extra electron added in the same
geometry as the neutral system. This gives a quantitatively better
gap estimate of $1.64$ eV.

The corresponding wave functions for these two states are shown in
figure \ref{fig:wf}. Surprisingly, they are different from a previous
calculation by Strohmaier {\it et al.} \cite{strohmaier} using 
semiempirical MNDO (modified neglect of
diatomic overlap). Their HOMO resembles our LUMO, while their LUMO has the
same symmetries as our HOMO although contributions from carbon atoms 
near the end of the 
pentacene molecule seem different. We do not know the reason for the 
discrepancy. However, we checked our result further by comparing it to an 
elementary H\"uckel-type calculation only including the 22 $p_z$ orbitals
each contributing one electron.
The signs of the wavefunction are the same as the ones obtained from the 
more elaborate SIESTA calculation.

Fig. \ref{fig:wf} (a)
shows the HOMO state. The main contribution to this wave function comes 
from the
$p_z$ orbital of the C atoms. In Fig. \ref{fig:wf} (b),
we show the coefficients of the
wave function expanded onto the first-zeta basis functions which resemble
most closely the atomic orbitals with $p_z$ symmetry. The positive values
are indicated by the filled circles and the negative values are shown
by the open circles. The large circles correspond to a magnitude between
0.12 and 0.3, while the small circles represent a value between 0.01 and
0.09. Both figures clearly demonstrate that the C atoms contributing to the
HOMO state are in a second neighbor (two large circles with a small circle
in between) configuration near the center but in the nearest neighbor
configuration at the two ends of the molecule.
It is important
to note that the relative phase of this molecular wave function varies
by $180^o$ in the nearest neighbor configurations at the ends of the molecule.
In Fig. \ref{fig:wf} (c), we plot the wave function of the LUMO state.
Similar to the HOMO state, the dominant contributing orbitals are also 
the $p_z$
orbitals. As shown in Fig. \ref{fig:wf} (d), similar second neighbor
configuration near the center as in HOMO state is observed but there are 
more nearest
neighbor configurations at the ends of the molecule. This difference at
the ends of the molecule explains why the HOMO state has lower energy
than the LUMO state because the latter state has a larger kinetic
energy due to rapid ($180^o$) phase changes.

\begin{figure}
\hspace{-2.5cm}
\begin{minipage}{8.5cm}
\includegraphics[width=11.0cm,angle=0]{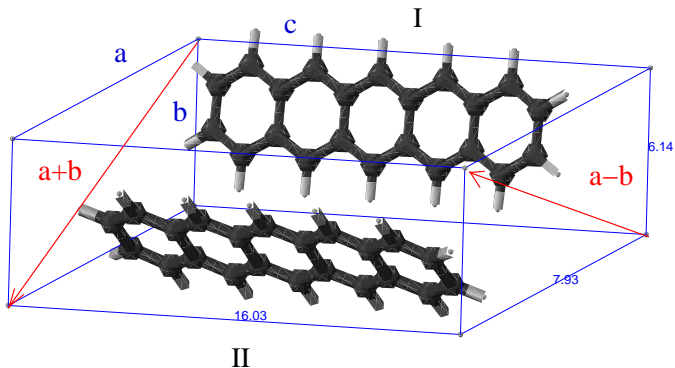}
\end{minipage}
\caption{\label{fig:UC61} Herringbone packing of two inequivalent pentacene
molecules in the unit cell of Campbell's model. The molecules are relaxed although three
atoms in each molecule are kept fixed. The Bravais lattice vectors {\bf a}, {\bf b}, and {\bf c}
are assigned, as well as the vectors {\bf a+b} and {\bf a-b}. Figure prepared with
the program MOLMOL\cite{molmol}.}
\end{figure}

\subsection{\label{subsec4.2}Molecular solid}

In figure \ref{fig:UC61} we show the structure for the molecular pentacene solid
according to the results of Campbell. The triclinic unit cell for each case is depicted
by thin solid lines. 
There are two molecules per unit cell labeled I and II following 
Campbell's notation which
are not parallel to each other, but are packed in a herringbone
fashion. 
The comparison of our atomic bond-lengths from 
the constrained relaxation with the experimental ones (in brackets) 
are shown in figure \ref{fig:pos} b). The centers of symmetry of molecule I
(0,0,0) and molecule II (1/2~,1/2~,0) in the space group P1 
are marked by arrows. The two redundant bond-lengths at the right end of 
molecule I illustrate this symmetry. The bond-lengths are qualitatively
correct and show the proper trends (alternating single-double bonds),
but are in general slightly larger than the experimental ones as typical for the DFT method.
Additionally, the experimental bond-lengths are not symmetric with respect to
the long molecular axes as opposed to the DFT result where the 
bonds are more symmetric similar to the single molecule (at least to
this precision). This reflects that noncovalent bonding by 
nonlocal interactions is not treated properly. This deficiency however
concerns mostly the total energy (and structure), less the single
particle wavefunctions (and bandstructure). 

\begin{figure}[t]
\hspace{-1.0cm}
\includegraphics[width=7.0cm,angle=-90]{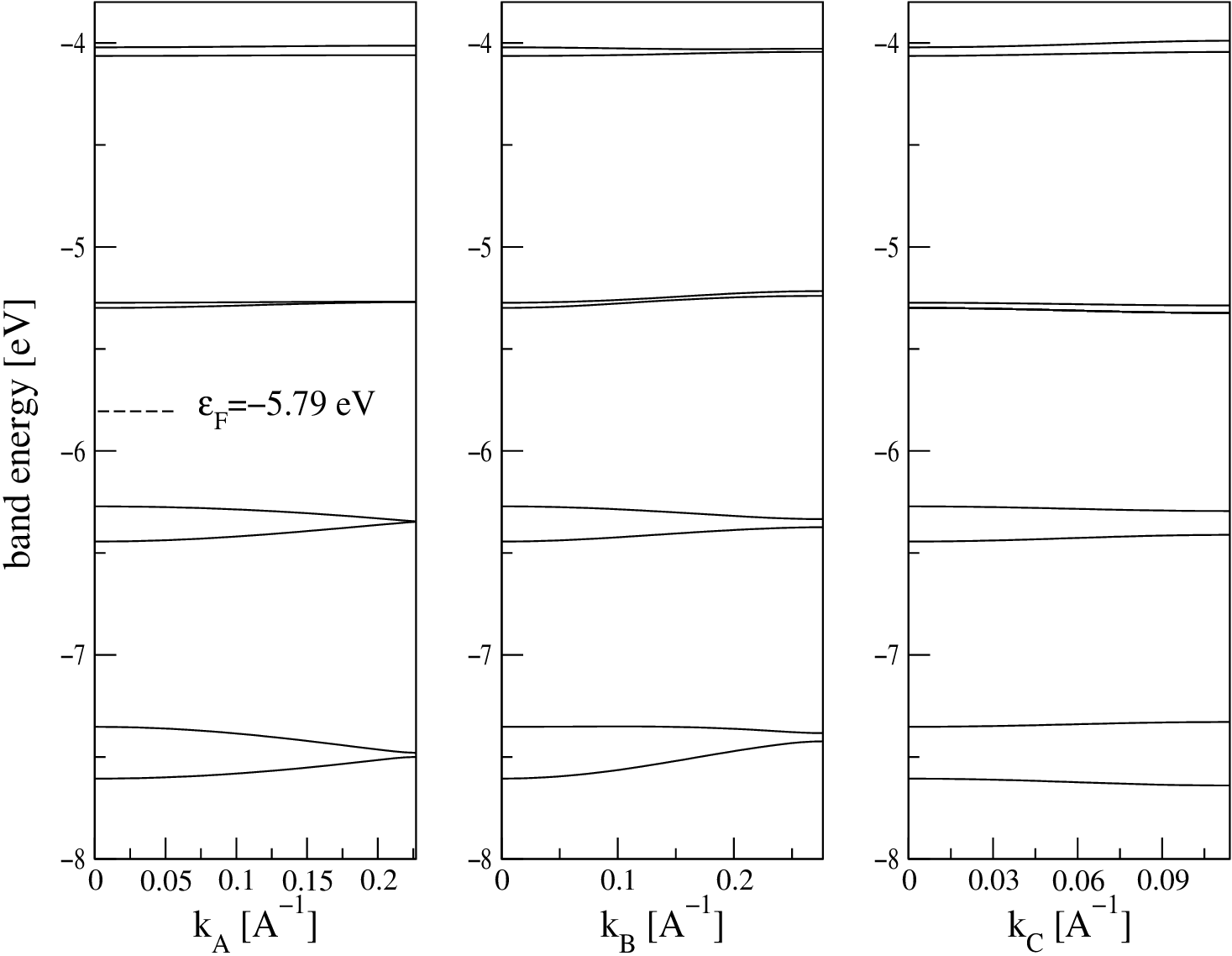}\\
\caption{\label{fig:bands61_1} Bandstructure along reciprocal
lattice vectors {\bf A}, {\bf B}, and {\bf C}. The LUMOs of the
two molecules per unit cell couple less than the HOMOs, also the
conduction band shows hardly any dispersion.}

\hspace{-0.9cm}
\includegraphics[width=7.0cm,angle=-90]{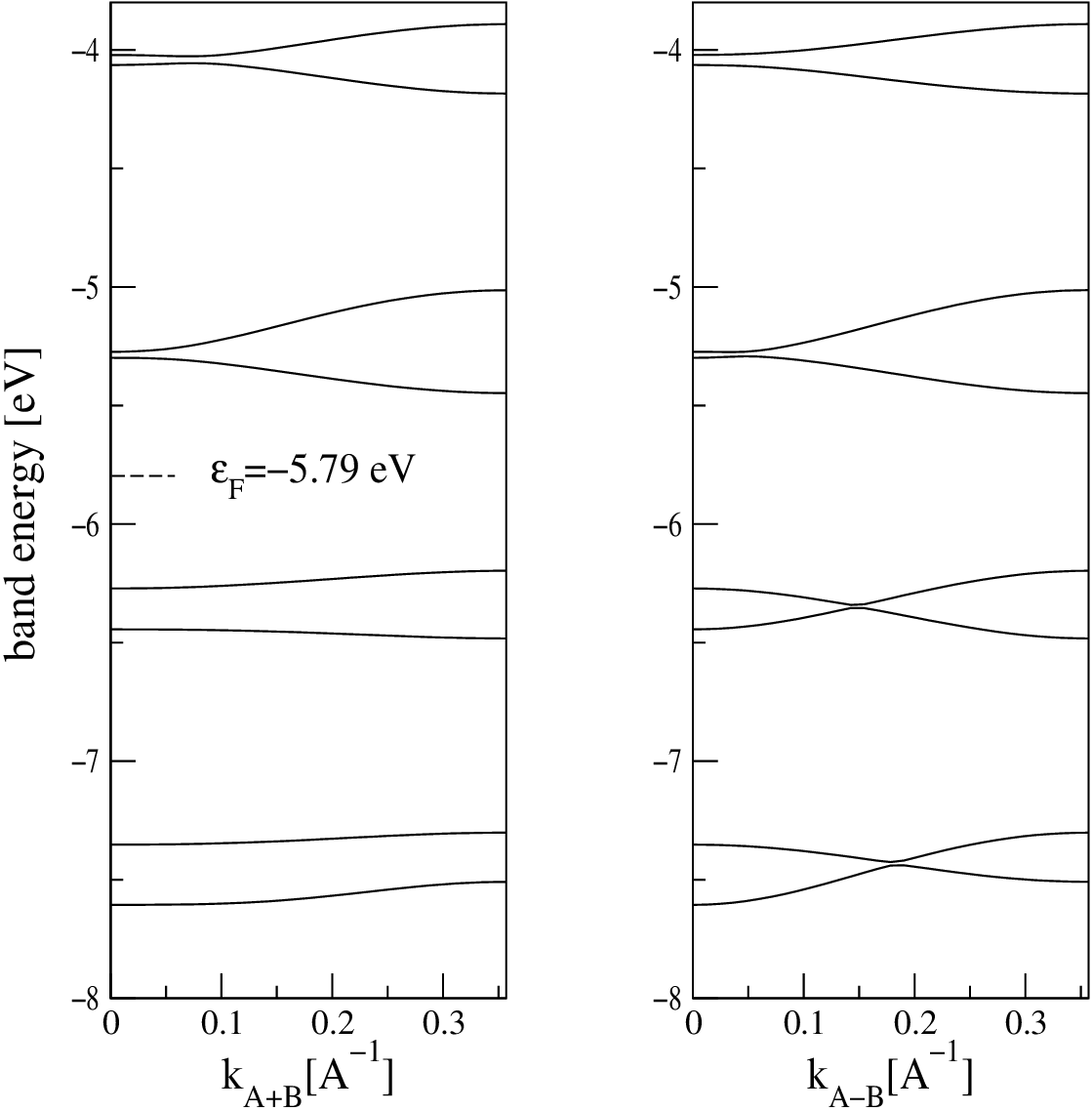}\\
\caption{\label{fig:bands61_2}Bandstructure along the stacking directions {\bf A+B} and {\bf A-B}.
 In these direction the bandwidths exceeds the polaron binding energy and band-like transport
 should be possible.}
\end{figure}

The band structure along the high symmetry axes for energies
ranging from -8 to -3.8 eV are depicted in figures
\ref{fig:bands61_1} and \ref{fig:bands61_2}. The bands always come
in sets of two, due to the near degeneracy of the states of two
pentacene molecules per unit cell. There are two important pieces
of information one can read off the band structure plots.

First, we want to estimate the transfer integrals t between the two molecules in a unit cell
and restrict ourselves to the couplings between the two HOMOs and between the two LUMOs.
These are most important for the hole and electron transport, respectively.
Since there is no momentum transfer involved for the intracell coupling, we can 
restrict ourselves to the $\Gamma$-point. At this point the bonding/antibonding splitting between 
two (identical) HOMO (LUMO) states is $2t_{H(L)}$. This gives 
$t_H=85$ meV for the HOMO coupling and $t_L=15$ meV for the LUMO coupling. 
Since the two molecules are in principle inequivalent 
where the level energy difference between the two HOMO (LUMO)
states is $\Delta E_{H(L)}$, the true coupling $t'_{H(L)}$ follows from
$2t'_{H(L)}=\sqrt{(2t_{H(L)})^2-\Delta E_{H(L)}^2}$. This equation of a coupled
inhomogeneous two level system was stated by Cornil {\it et al.}\cite{cornil} 
(their Eq. (1)). In their work an offset energy $\Delta E_{H(L)}$ as large as 61 meV (70 meV)
was quoted. While $\Delta E_{H}=61$ meV leads to $t'_H=80$ meV, $\Delta E_{L}=70$ meV gives 
a negative discriminant, i.e. our $t_L$ and Cornil's $\Delta E_{L}$ are not consistent. 
This at least illustrates that there cannot be a large offset in our case.


\begin{table}[b]
\begin{tabular}{c|cccc}
$[meV]$&VB-1&VB&CB&CB+1\\
\hline
{\bf A}&&73&29&\\
{\bf B}&&62&59&\\
{\bf C}&&23&25&\\
{\bf A+B}&41&75&149&260\\
{\bf A-B}&131&145&149&260
\end{tabular}
\caption{\label{tab:bw} Bandwidths of valence band (VB), the next lower band (VB-1),
 the conduction band (CB), and the next higher band (CB+1) in meV.}
\end{table}

Second, the dispersion and bandwidth reflect the coupling between
the unit cells and determines the crystal properties. The larger
the bandwidth the more delocalized the electronic states are at finite temperature
and the more one has transport by a band mechanism. If the bandwidth
becomes small and comparable to the polaron binding energy, then
excess charges are being self-trapped and need thermal activated
to migrate by a hopping mechanism.

The bandwidths are shown in table \ref{tab:bw}. At least along the
triclinic reciprocal lattice vectors {\bf A}, {\bf B}, and {\bf C}
they are all smaller than 73 meV and hence clearly smaller than
estimates of polaron binding energies ($\sim 200 meV$
\cite{brown}). This is likely to lead to a charge migration
mechanism of some sort of hopping in these directions. Either the
small polaron or the {\it multiple trapping and release} model
\cite{MTR} have been suggested. Our values for the band widths 
along {\bf A}, {\bf B}, and {\bf C} are in 
semi-quantitative agreement with previous
semiempirical ETH calculations by
Haddon and co-workers\cite{haddon}.

It was suggested by Cornil and co-workers \cite{cornil} using
extrapolation of results from finite clusters that the bandwidth
is much larger ($\sim 600$ meV) in the stacking directions
connecting the two inequivalent molecules. In their Fig. 1, {\bf
d}{${}_1$} connects the two molecules within same unit cell and
{\bf d}{${}_2$} between neighboring unit cells. Here we calculated
the bandstructure from first-principles along the high symmetry
directions {\bf A-B} and {\bf A+B} resembling Cornil's {\bf
d}{${}_1$} and {\bf d}{${}_2$} directions. We obtain that the
width of the CB and in particular the next higher band CB+1
increases drastically to $149$ and $260 meV$, respectively. On the
other hand, we only observe an increase of the width of the VB and
next lower band VB-1 in the {\bf A-B} direction. At a first
glance, they seem to be $211$ and $247$ meV. However, there is a
band crossing for the two highest occupied bands. Since the
triclinic unit cell has no symmetries (except the identity), the
crossing bands mix and open a $10$ meV gap (not resolved in plot). 
Due to the avoided crossing the widths of VB-1 and VB are only 131 and 145 meV,
respectively.

If one assumes a cross-over between polaron-like and band-like
transport at a bandwidth of 200 meV, band-like electron transport
should be possible in the stacking directions. However, we do not
observe bandwidths as large as those reported in Cornil's work
\cite{cornil}. When measuring the conductivity of a
pentacene crystal or extracting the mobility from transistor
characteristics, one should obtain thermally activated charge
hopping as well as temperature-independent band transport
behavior depending on the crystal orientation. 
Exactly this was observed by S. F. Nelson and co-workers
(see their Fig. 3) \cite{nelson}.

The question whether the width of the CB is smaller or equal to
the VB cannot be answered completely. Let us look at figure
\ref{fig:bands61_1} first. The LUMO splitting and dispersion of
the CB is clearly smaller than the HOMO splitting and dispersion
of the VB. This in line with conventional expectations. On the
other hand, figure \ref{fig:bands61_2} reveals that the bandwidths
of the CB along {\bf A+B} can indeed become larger than the VB
bandwidth. In particular, the widths of CB+1 is drastically
enlarged for {\bf A+B} and {\bf A-B}.

We further confirm by our method that there is hardly any
dispersion in the {\bf C} direction approximately along the
pentacene molecules. As already pointed out in Ref. \cite{cornil}
this leads to a quasi 2-dimensional character for charge
transport.

Finally, the fundamental band gap is 0.97 eV measured at the
$\Gamma$-point. This is slightly smaller than the Kohn-Sham
HOMO-LUMO gap for the single pentacene molecule, 1.10 eV, due to
band offsets from Brillouin zone folding and bonding/antibonding
splitting. As demonstrated in the case of a single pentacene
molecule the gap is drastically underestimated.

\section{\label{sec5}Summary} 
In conclusion, we used an ab-initio approach  
to determine the electronic properties and in particular the band-structure 
of a molecular solid of pentacene. A geometry optimization of the single
molecule yields an essentially planar molecule in agreement 
with experiment. The bond-lengths along the {\bf x}-direction (the direction of the 
molecule) do show $0.06 \AA$ difference between the bonds at the ends and the 
middle of the molecule. Both the HOMO and LUMO states originate from 
the $p_z$ orbitals on second neighbors C atoms in the center and on
nearest neighbors C atoms at the two ends of the molecule. The relative phases and the 
extent of overlapping between the neighboring $p_z$ orbitals differ for the 
two states. 

For the solid pentacene, calculations we carried out using the first-principles 
tight-binding code SIESTA for a pentacene molecular crystal with the experimentally determined
herringbone structure together with intramolecular distances resulting from a constrained
geomtry optimization. The resulting bond-lengths are in good agreement with experiment. 
The solid is predicted to be a large band-gap ($> 1.0 eV$)
semiconductor with an maximal bandwidth for electron transport of about $260$ meV
and a maximal bandwidth for hole transport of only $145$ meV.
It is found that the widths of the electronic bands depends strongly on the crystallographic direction.
Along the triclinic reciprocal lattice vectors the bandwidths are generally smaller than estimates based
on the small polaron binding energy whereas. Along the stacking directions a significantly larger
width is observed.
The present maximum bandwidths are much smaller than previous results obtained using 
the semiempirical INDO approach.
On the basis of the present theoretical results a band-like transport of charges in 
high-quality pentacene single
crystal with very high mobilities should be possible at low temperatures, as previously 
observed experimentally
for napthalene\cite{karl}.

{\bf Acknowledgement.} We would like to thank P. Ordej\'on, E.
Artacho, D. S\'anchez-Portal and J. M. Soler for providing us with
their \textit{ab initio} code SIESTA. RGE thanks support from
DOE-OS through BES-DMSE and OASCR-MICS under Contract No.
DE-AC0500OR22725 with UT-Battelle LLC. CYF thanks the support of
an NSF grant INT-987053, and NERSC at Lawrence Berkeley National
Laboratory. LHY is supported by the DOE under contract number
W-7405-ENG-48. CW and GW are funded by the Deutsche
Forschungsgemeinschaft (OFET Grant No. Wo 464/17-1)).



\begin{references}

\bibitem{horowitz} G. Horowitz, Adv. Mat. {\bf 10}, 365 (1998), and references therein;
                   J. Cornil, D. Beljonne, J. Ph. Calbert, and J. L. Br\'edas, Adv. Mater. {\bf 14},
                   1053 (2001).

\bibitem{nelson} S. F. Nelson, Y. -Y. Lin, D. J. Gundlach, and T. N. Jackson,
                 Appl. Phys. Lett. {\bf 72}, 1854 (1998).

\bibitem{lukas} S. Lukas, G. Witte, and Ch. Woll, Phys. Rev. Lett. {\bf 88}, 028301 (2002).

\bibitem{heeger} A. J. Heeger, in {\it Highly conducting one-dimensional solids};
                 J. T. Devreese, R. P. Evrard, V. E. van Doren, Eds. (Penum Press, New York, 1979);
                 p. 69 and references therein


\bibitem{holstein} T. Holstein, Ann. Phys. (NY) {\bf 8},343 (1959).

\bibitem{brown} A. R. Brown {\it et al}, Synth. Met. {\bf 88}, 37 (1997).

\bibitem{cornil} J. Cornil, J. Ph. Calbert, and J. L. Br\'edas, J. Am. Chem. Soc. {\bf 123}, 1250 (2001);
                 J. L. Br\'edas {\it et al.}, Synth. Met. {\bf 125}, 107 (2002).

\bibitem{haddon} R. C. Haddon {\it et al.}, J. Phys. Chem B {\bf 106}, 8288 (2002).

\bibitem{siesta_code} http://www.uam.es/siesta

\bibitem{siesta_papers} D. S\'anchez-Portal, P. Ordej\'on, E. Artacho,
         and J. M. Soler, Int. J. Quantum Chem. {\bf 65}, 453 (1997);
         E. Artacho, D. S\'anchez-Portal, P. Ordej\'on,
         A. Garcia, and J. M. Soler, Phys. Status solidi (b)
         {\bf 215}, 809 (1999); P. Ordej\'on, E. Artacho, and
         J. M. Soler, Phys. Rev. B {\bf 53}, R10441 (1996).

\bibitem{TM} N. Troullier and J. L. Martins, Phys. Rev. B {\bf 43}, 1993 (1991).

\bibitem{KB} L. Kleinman and D. M. Bylander, Phys. Rev. Lett. {\bf 48}, 1425 (1982).

\bibitem{sankey} O. F. Sankey and D. J. Niklewski, Phys. Rev. B {\bf 40},
                 3979 (1989).

\bibitem{gaussian} S. Huzinaga, J. Andzelm, {\it et al.} ed ``Gaussian basis
                   sets for molecular calculations'',
                   Elservier Science Pub. Co., New York, (1984).

\bibitem{PBE} J. P. Perdew, K. Burke, and M. Ernzerhof, Phys. Rev. Lett. {\bf 77},
              3865 (1996).

\bibitem{campbell} R. B. Campbell, J. Monteath Robertson and J. Trotter,
                   Acta Cryst. {\bf 14}, 705 (1961).

\bibitem{holmes} D. Holmes, S. Kumaraswamy, A. J. Matzger, and
         K. Peter C. Vollhardt, J. Euro. Chem. {\bf 5}, 3399 (1999).

\bibitem{siegrist} Th. Siegrist {\it et al.}, Angew. Chem. Int. Ed. {\bf 40}, 1732 (2001).

\bibitem{adamo} C. Adamo and V. Barone, J. Chem. Phys. {\bf 108}, 664 (1998).

\bibitem{kafafi} S. A. Kafafi, J. Phys. Chem. A {\bf 102}, 10404 (1998).

\bibitem{park} S. P. Park, S. S. Kim, J. H. Kim, C. N. Whang, and S. Im,
               Appl. Phys. Lett. {\bf 80}, 2872 (2002).

\bibitem{gap_probl} W. Hanke and L. J. Sham, Phys. Rev. Lett. {\bf 43}, 387 (1979);
                    M. S. Hybertsen and S. G. Louie, Phys. Rev. Lett. {\bf 55}, 1418 (1985).


\bibitem{SCF} R. O. Jones and O. Gunnarsson, Rev. Mod. Phys. B {\bf 13}, 4274 (1989);
C. J. Cramer and F. J. Dulles, J. Am. Chem. Soc. {\bf 116}, 9787 (1994).


\bibitem{gopenmol} L. Laaksonen, J. Mol. Graph. {\bf 10}, 33 (1992);
 D. L. Bergman, L. Laaksonen, and A. Laaksonen, J. Mol. Graph. {\bf 15}, 301 (1997).
gOpenMol: A graphics program for the anaylsis and display of molecular dynamics trajectories.

\bibitem{strohmaier} R. Strohmaier, J. Petersen, B. Gompf, and W. Eisenmenger, Surf. Sci. {\bf 418}, 91 (1998).

\bibitem{molmol} R. Koradi, M. Billeter, and K. W\"uthrich, K., J. Mol. Graph. {\bf 14}, 51 (1996).
       MOLMOL: a program for display and analysis of macromolecular structures

\bibitem{MTR} P. G. Le Comber, W. E. Spear, Phys. Rev. Lett. {\bf 25}, 509 (1970).

\bibitem{karl} W. Warta, R. Stehle, N. Karl, Appl. Phys. A {\bf 36}, 163 (1985); N.Karl: 
"Charge-Carrier Mobility in Organic Crystals",
in: Organic Electronic Materials, R. Farchioni and G. Grosso (eds.), 
Part II: Low Molecular Weight Organic Solids (chapter 8), (Springer Verlag, Berlin 2001) 


\end{references}
\end{document}